\documentstyle[aps,twocolumn,times,epsf]{revtex}
\def\void{}
\def\labelname{}
\def\labelmark{}

\newenvironment{formula}[1]%
{\def\labelname{#1}
\ifx\void\labelname\begin{displaymath}
\else\labelmark\begin{equation}\label{\labelname}\fi}%
{\ifx\void\labelname\end{displaymath}\else\end{equation}\fi}

\newenvironment{formulas}[1]%
{\def\labelname{#1}
\ifx\void\labelname\begin{displaymath}
\else\labelmark\begin{equation}\label{\labelname}\left.\fi
\begin{array}{lllllll}}%
{\end{array}\ifx\void\labelname\end{displaymath}\else\qquad\right\}
\end{equation}\fi}

\newcommand{\Fig}[1]{Fig.~\ref{#1}}
\newcommand{\bort}[1]{}
\def\.{~~.}

\def\simleq{\; \raise0.3ex\hbox{$<$\kern-0.75em
      \raise-1.1ex\hbox{$\sim$}}\; }
\def\simgeq{\; \raise0.3ex\hbox{$>$\kern-0.75em
      \raise-1.1ex\hbox{$\sim$}}\; }

\def\noi{\noindent}
\def\ie{{\it i.e.\/}}

\def\({\left(}
\def\){\right)}

\def\>{\rangle}
\def\<{\langle}

\def\Tr{{\rm Tr}}
\def\T{\overline{T}}
\def\p{\overline{p}}
\def\P{{\cal P}}
\def\n{{\hat n}}

\def\bfor#1{\begin{formula}{#1}}
\def\efor{\end{formula}}

\def\sec{{\rm s}}

\FR
\begin{document}
\title{Atomic Beam Correlations and the Quantum State of the  Micromaser}

\author
{Per Elmfors\thanks{email: elmfors@cern.ch}$~~^{a)}$,
Benny Lautrup\thanks{email: lautrup@connect.nbi.dk}$~~^{a,b)}$,
and  Bo-Sture Skagerstam \thanks{email: tfebss@fy.chalmers.se}
$~~^{a,c)}$\\[3mm]
\small
$^{a)}$ CERN,   TH-Division, CH-1211 Geneva 23, Switzerland \\[1mm]
\small $^{b)}${\sc connect}, The Niels Bohr Institute, Blegdamsvej 17,\\
\small        DK-2100 Copenhagen, Denmark \\[1mm]
\small $^{c)}$Institute of Theoretical Physics, Chalmers University of
 Technology, \\
\small G\"oteborg University, S-412 96 G\"oteborg, Sweden, \\
and Department of Physics, University of Trondheim, N-7055 Dragvoll, Norway
}

\author{\begin{quotation}
\small Correlation measurements on the states of two-level atoms having  passed
through a micromaser at different times can be used to infer properties of  the
quantum state of the radiation field in  the  cavity.  Long(short)  correlation
length in time is to some extent associated with  super(sub)-Poissonian  photon
statistics. The correlation length is also an indicator of  a  phase  structure
much richer than what is revealed by the usual  single-time  observables,  like
the atomic inversion or the Mandel quality factor.  In  realistic  experimental
situations the correlations may extend over many times the decay  time  of  the
cavity. Our assertions are verified by comparing theoretical calculations  with
a high-precision Monte-Carlo simulation of the micromaser system.
\end{quotation}
}

\maketitle

\newpage

\normalsize


The          one-photon          atomic          transition          micromaser
\cite{Haroche83,Meschede85,Walther88,Walther95}  provides  an   impressive
experimental
realization of the interaction between a two-level atom and a second  quantized
single-mode  electromagnetic  field.  The   microlaser   \cite{An94}   is   the
counterpart in the optical regime. Here we focus our interest on the micromaser
system for which quantum collapse  and  signs  of  quantum  revival  have  been
observed \cite{Rempe87}. A superconducting niobium cavity,  cooled  down  to  a
temperature of $T=0.5\,{\rm K}$ (corresponding to a thermal  photon  occupation
number of $n_b=0.15$ at the maser frequency of 21.5  GHz),  has  been  used  to
study the quantum state of the radiation field \cite{Rempe90,Rempe91}. The high
quality factor of the cavity  corresponds  to  a  photon  lifetime  of  $T_{\rm
cav}=0.2~{\rm s}$.

In this context, a basic physical feature is the close connection  between  the
cavity's steady-state photon statistics and the fluctuations in the  number  of
atoms in the lower maser level, for a fixed cavity transit time $\tau$  of  the
atoms \cite{Filipowicz86,Rempe90a}. The experimental results of  \cite{Rempe90}
are clearly consistent with the  appearance  of  non-classical,  sub-Poissonian
statistics of the radiation field, and exhibit the intricate relation  between
the atomic beam and the quantum state of the cavity.

In this letter we study in more detail this relation, and discuss the role of
correlations for revealing the quantum state of the  micromaser  system.  The
sequence of outgoing atoms is viewed as a one-dimensional binary spin  chain,
each spin representing the state of an atom after the  interaction  with  the
cavity. The spin average is closely related to what  is  usually  called  the
atomic inversion. From the second-order correlation  functions  of  the  spin
chain we are immediately led to the physical concept of an atomic correlation
length $\xi_A$, characterizing the long-range correlations in time  in  the
outgoing atomic beam.

By means of Monte Carlo simulations we study the dynamical generation  of  such
correlations in the atomic beam and how these would  show up in  an
actual experimental situation. We discuss the  set  of  characteristic  dynamic
phases revealed by $\xi_A$, and the implications for the underlying  photon
distribution in the cavity.



The theory of the micromaser has been developed in \cite{Filipowicz86}, and  we
follow the notation of that  paper.  In  the  experiments,  excited  atoms  are
injected into the cavity at a rate $R$ high enough to pump up the  cavity  from
vacuum, \ie\ $R>\gamma=1/T_{\rm cav}$, or $N>1$ in terms of  the  dimensionless
flux variable $N=R/\gamma$. The atom-field interaction time $\tau$ is  so  much
shorter than the average time between two atoms, $\T=1/R$,  that  with  a  very
high  probability  only  one  atom  at  a  time  is  found  inside  the  cavity
\cite{Wehner94}. Since $\tau$ is also much shorter than $T_{\rm cav}$,  damping
effects may be ignored while the atom passes through the  cavity.  If  an  atom
enters at time $t$ and interacts  for  a  time  $\tau$,  then  the  statistical
operator       of       the        whole        system        changes        to
$\rho(t+\tau)=e^{-iH\tau}\rho(t)e^{iH\tau}$, where $H$ is the total Hamiltonian
of the atom-field interaction. It can be approximated by  the  Jaynes--Cummings
(JC) Hamiltonian \cite{Jaynes63}. This Hamiltonian has  the  property  that  it
only affects the reduced density operator $\rho_F(t)=\Tr _{A}(\rho(t))$ of  the
radiation field along its diagonals, so that if  $\rho_F(t)$  is  diagonal  ---
which   we    assume    ---    then    so    is    $\rho_F(t+\tau)$.    Writing
$\rho_F(t)=\sum_{n=0}^\infty  p_n(t)|n\>\<n|$  we   may   express   the   above
interaction as a linear operator  acting  on  the  infinite-dimensional  vector
$p(t)=\{p_0(t),p_1(t),\ldots\}$,   \ie\   as   $p(t+\tau)=M(\tau)p(t)$,   where
$M=M^++M^-$ is composed of two  parts,  representing  either  that  the  photon
occupation number is unchanged (the atom is  in  state  $+$),  or  that  it  is
increased by 1 due to the decay of the atom (the atom is in  state  $-$).  From
the JC--model we have

\begin{formulas}{defMplus/minus}
M^+_{nm}(\tau)=(1-q_{n+1}(\tau))\delta_{nm}~~,~~~~~~~~~~~~~~~~~~~~~~~~~~\\
M^-_{nm}(\tau)=q_n(\tau)\delta_{n-1,m}~~,     ~~~~~~~~~~~~~~~~~~~~~~~~~~\\
\end{formulas}

\noi where at resonance between the  cavity  mode  and  the  atomic  transition
$q_n(\tau)=\sin^2(\Omega\tau\sqrt{n})$. The   quantity
$\Omega$ is the single-photon Rabi frequency \cite{Jaynes63}.

Let the next atom arrive in the cavity after a  time  $T\gg\tau$.  During  this
interval the cavity damping is described by  a  conventional  master  equation,
which also preserves the diagonal form of the cavity density matrix. It may  be
brought to the form $\stackrel{.}{p}(t)=-\gamma L p(t)$, where $\gamma$ is  the
characteristic damping constant of the cavity and $L$ is the matrix:

\begin{eqnarray}
L_{nm}=&(n_b+1)(n\delta_{n,m}-(n+1)\delta_{n+1,m})\nonumber\\
&+n_b((n+1)\delta_{n,m}-n\delta_{n-1,m})\.
\end{eqnarray}

\noi The statistical state of the cavity when the next  atom  arrives  is  thus
given by

\begin{formula}{stat1}
p(t+T)=e^{-\gamma LT}M(\tau)p(t)\.
\end{formula}

\noi The time intervals between atoms are assumed to  be  Poisson-distributed
$d\P(T)=\exp(-RT)RdT$ with an average interval $\T=1/R$ (for a discussion  of
non-Poissonian beam statistics see \cite{Briegel95}), and consequently we  may
average the exponential of Eq.~(\ref{stat1}) to get

\begin{formula}{aveL}
p(t+\T)\simeq \frac1{1+\gamma\T L}M(\tau) p(t)\equiv S(\tau,\T)p(t)\.
\end{formula}

\noi    The    equation    for    statistical    equilibrium    thus    becomes
$M(\tau)\p=(1+\gamma\T      L)\p$,      which      has       the       solution
\cite{Filipowicz86,Lugiato87} for $n\ge1$

\begin{formula}{defpstar}
\p_n=\p_0\prod_{m=1}^n
\frac{n_b}{n_b+1}\(1+\frac{q_{m}}{\gamma\T n_b m}\)\.
 \end{formula}

\noi The overall constant $\p_0$ is determined by $\sum_{n=0}^\infty \p_n=1$.


Let the state of an atom emerging from the cavity be characterized by a  binary
``spin'' variable $s=\pm1$ where $+1$ denotes the excited level. In statistical
equilibrium the probability $\P(s)$ of finding the atom in the state $s$  after
the interaction is given by $\P(s)=u^\top M^s\p$ where $u$ is a vector with all
entries equal to one, representing the trace over the cavity. If the  detection
efficiency   is   not   100\%   these   formulas   are   modified   accordingly
\cite{Briegel94}. The average spin is $\mu=\<s\>=\P(+)-\P(-)$ (and  the  atomic
inversion $-\mu/2$). Its variance is $\sigma^2=\<(s-\mu)^2\>=1-\mu^2$.  From
the spin average we  may  in  statistical  equilibrium  determine  the  average
occupation number of the cavity mode $\<\n\>=n_b+(1-\mu)/2\gamma\T$.

The joint probability $\P_k(s_1,s_2)$ for observing the states  of  two  atoms,
with $k$ unobserved atoms between them, is given by

\begin{formula}{defjointp}
\P_k(s_1,s_2)=u^\top M^{s_2}S^k \frac1{1+\gamma\T L}M^{s_1}\p~~,
\end{formula}

\noi where the matrix $S=S(\tau,\T)$ is defined in Eq.~(\ref{aveL}). From  this
we derive  the  $k$-step  correlation  function  $\<ss\>_k=\sum_{s_1,s_2}s_1s_2
\P_k(s_1,s_2)= \P_k(+,+)+ \P_k(-,-)- \P_k(+,-)- \P_k(-,+)$. It is easy to  show
that $\P_k(+,-)=\P_k(-,+)$ using  the  stationarity  condition  for  $\p$.  The
normalized atomic correlation is $\gamma_k^A = (\<ss\>_k-\<s\>^2)/\sigma^2$.

The state of the cavity can be characterized by the  average  $\<\n\>$  of  the
occupation number operator $\n$ and higher-order observables such as the Mandel
quality factor $Q_f= \<(\n-\<\n\>)^2\>/\<\n\>-1$, which  has  the  property  of
being negative for sub-Poissonian  (and  non-classical)  states  of  the  field
\cite{Mandel79}. We may further characterize the state of the  radiation  field
by correlations between cavity observables at different times. If  for  example
the occupation number $\n$ is measured twice, with $k$ unobserved atoms passing
in between, we consider the correlation function

\begin{formula}{defphotoncorr}
\<\n\n\>_k=u^\top \n S^k \n \p\.
\end{formula}

\noi   As   before   we   define   the    normalized    correlation    to    be
$\gamma^F_k=(\<\n\n\>_k-\<\n\>^2)/\<(\n-\<\n\>)^2\>$.

If $k$ is sufficiently large we define the asymptotic correlation lengths  $\xi
_{A}$ and $\xi _{F}$ by

\begin{formula}{defcorrelations}
\gamma_k^{A,F} \simeq \gamma^{A,F}\exp \(-\frac{k}{R\xi _{A,F}}\)~~,
\end{formula}

\noi which  isolates  the  $k$-dependence.  The  $R$-factor  secures  that  the
correlation lengths are measured  in  units  of  physical  time.  That
$\xi\equiv
\xi_A=\xi_F$ follows from the fact that the time evolution  is  governed  by
the same matrix in both Eqs.~(\ref{defjointp})  and  (\ref{defphotoncorr}).
The correlation length is therefore a convenient probe of the dynamics
of the micromaser system.

The generation of a sequence of outgoing atoms is a Markov process  defined  by
the matrix $S = (1+\gamma\T L )^{-1}M(\tau)$, and correlations over long  times
are governed by its eigenvalues $\lambda_n$. They can  be  shown  to  be  real,
non-degenerate  and  bounded  by   $1=\lambda_{0}>   \lambda_{1}   >...\ge   0$
\cite{ElmforsLS95}. Denoting the corresponding  right  (left)  eigenvectors  by
$p^{(n)}$ ($u^{(n)}$) we can decompose $S^k$ as

\begin{formula}{Rdecomp}
        S^k= p^{(0)}\otimes u^{(0)} + \sum_{n=1}^\infty
(\lambda_n)^kp^{(n)}\otimes u^{(n)}~~.
\end{formula}

\noi The contribution to $\gamma_k^A$ or $\gamma_k^F$  from  the  $\lambda_0=1$
component of $S^k$  cancels  out  and  it  is  the  next-to-leading  eigenvalue
$\lambda_1$   that   determines   the   correlation   length   through    $R\xi
=-1/\ln\lambda_1$.

\begin{figure}[htp]
\unitlength=0.5mm
\begin{picture}(160,140)(0,0)
\includegraphics{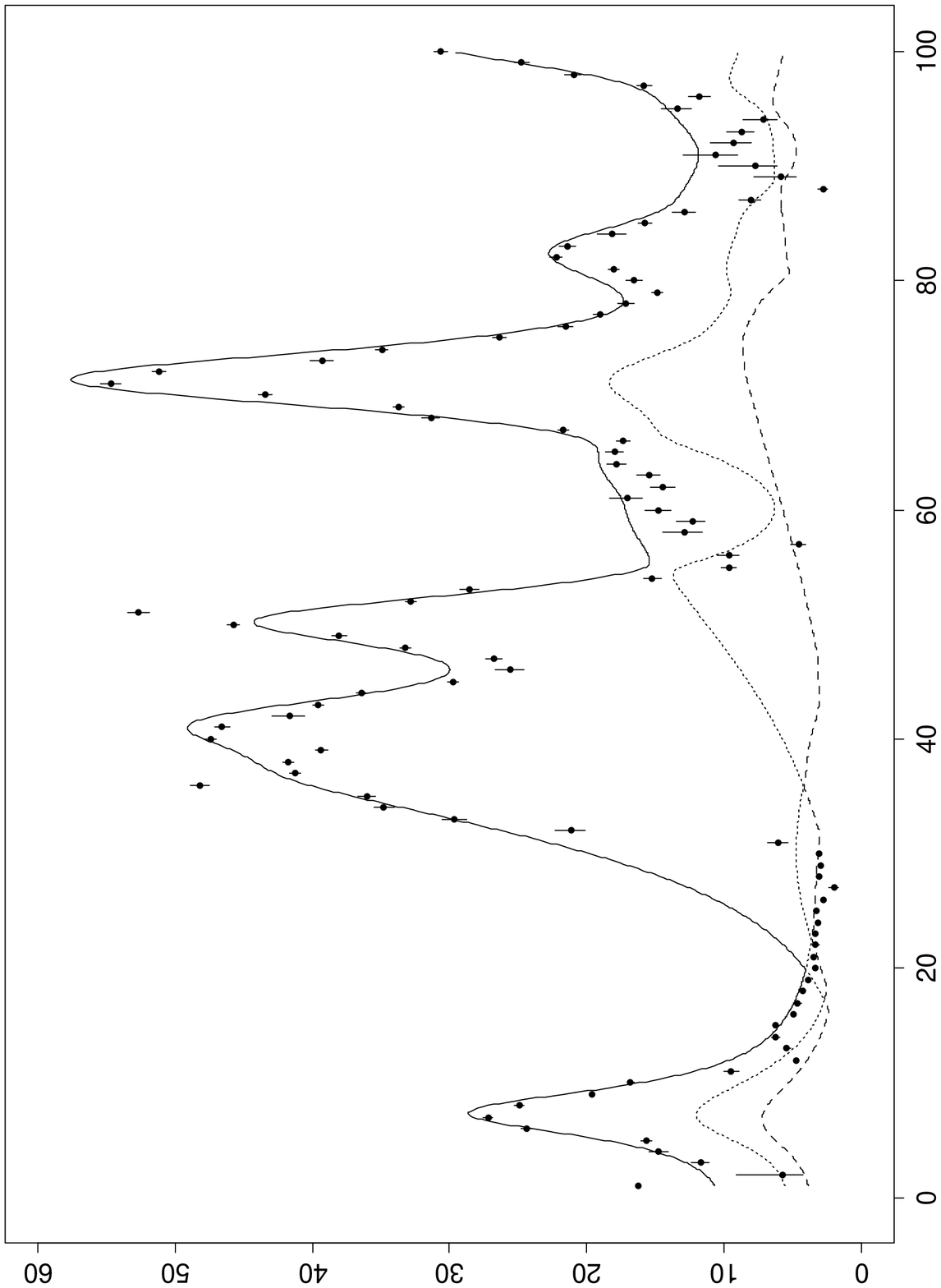}
\put(75,5){\small$\tau~[\mu {\rm s}]$}
\put(-15,73){\small$R\xi$}
\put(10,125){\small$^{85}$Rb 63p$_{3/2}\leftrightarrow$ 61d$_{5/2}$}
\put(10,115){\small$R=50~{\rm s}^{-1}$}
\put(10,105){\small$n_b=0.15$}
\put(10,95){\small$\gamma=5~{\rm s}^{-1}$}
\end{picture}
\caption[]{\protect\small Comparison of theory (solid curve) and MC data (dots)
for the correlation length $R\xi$. The dotted and dashed curves correspond  to
subleading eigenvalues ($\lambda_{2,3}$) of the matrix $S$. The parameters  are
those of the experiment in \cite{Rempe90}. }
\label{xi}
\end{figure}

In order to determine $\xi$ we have used two methods, a Monte Carlo  simulation
of the dynamical process and a direct numerical calculation of $\lambda_n$.  We
consider the maser  transition  $63  p_{3/2}  \leftrightarrow  61  d_{5/2}$  of
$\,^{85}\mbox{Rb}$ with the single-photon Rabi frequency $\Omega = 44$ kHz.  We
choose the rate $R$ to be 50 atoms/s in order to be close to  the  experimental
situation of \cite{Rempe90}. In the MC simulation  we  used  a  sample  of
$10^6$ atoms for  100  different  values  of  $\tau$  (the  corresponding  real
experiment would take a little less than a month to perform non-stop with 100\%
measuring efficiency!). We  have  compared  the  MC-data  and  the  theoretical
average of the atomic inversion as well as the Mandel quality factor. The
agreement  between  theory  and  the  numerical  experiment  is  excellent  and
constitutes a consistency check of our computational  method.  The  correlation
length is measured by a least-squares fit to the data for $1<k<K$, where $K$ is
determined by the onset of noise which  almost  universally  happens  when  the
correlation has fallen to about 0.1\%. Typical values of $K$ range from  10  to
250, depending on whether the correlation quickly  or  slowly  drops  into  the
noise. This leads to quite different reliability levels for the  extraction  of
the values of the correlation length. Our results are presented in \Fig{xi}.
The
error bars on the MC data represent the precision of the fit to the exponential
form Eq.~(\ref{defcorrelations}), but do not include the systematic errors  due
to the difficulty in reaching the asymptotic region. %

The other numerical method we have used is to find the roots  of  the  equation
$\det[M-\lambda(1+\gamma\T  L)]=0$,  by  truncating  the   matrix   to   finite
dimension. These roots are the eigenvalues of $S=(1+\gamma\T L)^{-1}M$. Writing
the above equation as an eigenvalue problem for the matrix  $M-\alpha  L$  with
$\alpha=\lambda\gamma\T$, we recognize that it is a Jacobi matrix (with
non-vanishing
elements only on the main diagonal and its two neighbouring  subdiagonals).
The
off-diagonal elements are strictly positive for non-negative real $\alpha$.  It
is known \cite{Fritz79} that such a matrix (for fixed $\alpha$) has  only  real
and non-degenerate eigenvalues. We have shown that the eigenvalues of $S$  also
have these properties, and that $\lambda_n\leq1$, which  is  required  for  the
convergence of the stochastic process. Numerical  calculations  of  $\lambda_n$
converge rapidly for matrices of dimension above 50. They also agree well  with
the MC simulation, as can be seen in \Fig{xi}. For  $\tau\simeq  20-30~\mu{\rm
s}$ the MC result predicts shorter correlation  lengths  than  the  theoretical
calculation, but we can see that they agree with a  subleading  eigenvalue.  We
surmise that the reason for  this  is  that  contributions  of  the  subleading
components of $S$ are large for those values of $k$ that can be reached in  the
MC calculation. In order to extract the actual long-time correlation length  in
this parameter region, it would be necessary to  go  much  larger  $k$  and  to
increase the statistics considerably. Similar difficulties are expected  to  be
encountered in a real experiment.

\begin{figure}[htp]
\unitlength=0.5mm
\begin{picture}(160,135)(0,0)
\includegraphics{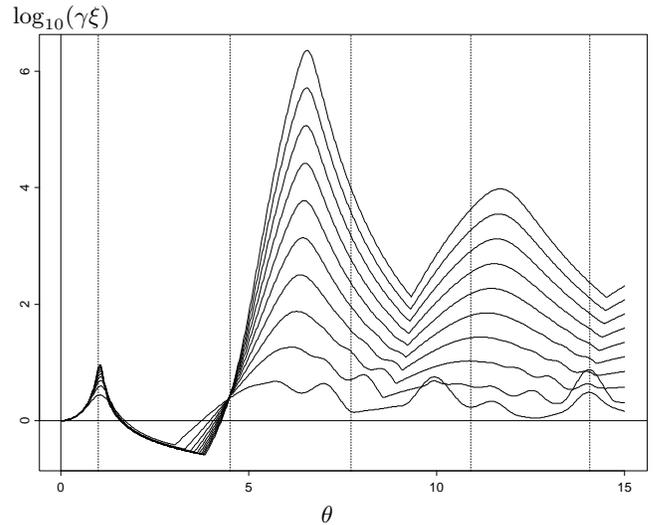}
\put(77,2){\small$\theta$}
\put(-5,135){\small$\log_{10}(\gamma\xi)$}
\end{picture}
\caption[]{\protect\small  The  logarithm  of  the   correlation   length
(in   units   of   $1/\gamma=N/R$)   for   $n_b=0.15$   as   a   function    of
$\theta=g\tau\sqrt{N}$ for various values of $N=10,20,\ldots,100$. The vertical
lines indicate where new local  maxima  in  the  photon  distribution  function
appear.  Notice  that  for  $\theta>\theta_1\simeq  4.494$,  where  the  photon
distribution acquires two maxima, the logarithm of the correlation length grows
linearly with $N$ for large $N$. }

\label{logxi}
\end{figure}

The  example  of  the  correlation  lengths  shown  in  \Fig{xi}  have  typical
parameters  of  the  damping   and   the   flux   corresponding   to   a   real
experiment~\cite{Rempe90}. The amount of structure found in this figure  is  at
first glance rather surprising. The  mean  photon  number  shows  comparatively
little structure in this region. We shall now explain the origin  of  structure
in this figure, and refer to \Fig{logxi}, when  necessary,  which  depicts  the
logarithm of the correlation length for larger flux values. The natural scaling
variable in the large flux limit is $\theta=g\tau\sqrt{N}$  which  is  the  one
used in \Fig{logxi}. Details of the theoretical derivations will  be  presented
in \cite{ElmforsLS95}.

In the first part of \Fig{xi}, for $0<\tau<20\,\mu s$, the  correlation  length
shows a peak around $\tau=7.2\,\mu s$ which corresponds to the maser transition
at $\theta=1$. For smaller values of $\tau$ the field of the cavity has thermal
statistics  with  average  number  of  photons  independent  of  the  flux   $N
(=R/\gamma)$ in the large flux limit.  Above  the  peak,  the  average  photon
number grows proportionally with the flux. This peak,  which  is  also  clearly
present  in  $\mu$  and  $Q_f$,  is   well   described   by   a   semiclassical
approximation~\cite{Filipowicz86,Guzman89}. In \Fig{logxi} we see that in  this
region the correlation length remains constant in the large flux  limit  except
at the maser transition point where it actually grows like $\sqrt{N}$.

At $\tau\simeq20\,\mu s$,  there  seems  to  be  crossings  of  eigenvalues  in
\Fig{xi}. Looking closer at these points, we  find  that  there  is  no  actual
crossing, consistent  with  the  result  mentioned  above  that  there  are  no
degenerate eigenvalues. On the other hand, the corresponding  eigenvectors,  in
this case $p^{(1)}$ and $p^{(2)}$, do cross in the sense that  the  eigenvector
corresponding to $\lambda_2$ after the crossing  is  very  close  to  $p^{(1)}$
before the crossing. This has no analogue  in  $\mu$  or  $Q_f$.  From  the  MC
simulation we also see that the system may remain  dominated  by  a  subleading
eigenstate as $\tau$ passes through the transition point. In \Fig{logxi} we see
how this crossing point as the flux increases creeps closer to  $\theta_1\simeq
4.494$,  where  the  photon  distribution  function  gets  a   second   maximum
\cite{ElmforsLS95}.

Above $\tau\simeq 32\,\mu s$ the correlation length increases dramatically. The
reason for that can be traced back to the appearance of two local maxima in the
photon distribution function, and the tunneling time between them, as indicated
in \cite{Filipowicz86} and calculated explicitly in \cite{ElmforsLS95}. We have
verified that, as the flux $R$ of atoms  increases,  the  maximal  correlation
length grows exponentially in this region.  For  a  value  of  $R=200/{\rm  s}$
(which is merely 4 times the flux in the  experiment  of  \cite{Rempe90}),
the correlation length at the leading peak extends to 300 times the decay  time
of the cavity (\ie\ a whole minute!). Such a violent behaviour is not
reflected  in
$\mu$ or $Q_f$~\cite{Filipowicz86,Guzman89}.  For  these  quantities  the  most
dramatic behaviour is a finite jump in $\mu$ and a narrow peak of finite height
in $Q_f$ at the second maser transition, which coincides with  the  exponential
peak in $\xi$.

The second maser transition occurs when there is a jump in the position of  the
global maximum of the photon distribution function. For $N=10$ this happens  at
$\tau\simeq51.7\mu\sec$. The three high peaks  at  $\tau$  around  40,  50  and
70$\,\mu s$ have, however, a different explanation. They  are  related  to  the
well-known {\em trapping states} \cite{Meystre88} which occur when  $q_n(\tau)$
vanishes. This happens for $\tau\sqrt{n}=k\pi$ with integer $k$. For  vanishing
$n_b$,   this   effect   truncates   the   equilibrium   photon    distribution
Eq.~(\ref{defpstar}) at a definite photon number.  Finite  $n_b$  smoothes  out
this effect, but it is still amply visible in \Fig{xi}. The  dominant  trapping
peaks are numerically given by $\tau=41.2\,\mu\sec,~ 50.5\,\mu\sec,~  71.4\,\mu
\sec$ which agrees very well with \Fig{xi}. In \Fig{logxi}  we  see  how  these
peaks are rapidly suppressed relative to the leading and exponentially  growing
tunneling peak for larger flux. The trapping peaks occur at constant  positions
in $\tau$ but move away as $\sqrt N$ in $\theta$, whereas  the  tunneling  peak
remains at fixed $\theta$.

In the semiclassical regime there is a  clear  connection  between  the  long
correlation at the maser transition at $\theta=1$  and  the  large  value  of
$Q_f$, i.e. super-Poissonian photon statistics. For
$\theta>\theta_1\simeq4.494$,  and
for large flux, $Q_f$ likewise has peaks at the same  positions  as  $\xi$,
namely when there are jumps in the position of  the  global  maximum  of  the
photon distribution. In this sense there is a clear relation between  $\xi$
and photon statistics in the cavity. On the other  hand,  at  $\theta_1$  for
large flux the Mandel factor remains negative while $\xi$  starts  growing
exponentially, and this growth reveals the appearance  of  the  second  local
maximum in the photon distribution which is not reflected in  the  single-time
observables.


B.L. and B.-S.S. wish to thank Gabriele Veneziano and the TH-Division  for  the
hospitality at CERN while this work was carried out. The authors also  wish  to
thank I. Lindgren for providing a guide to the experimental  work  and  T.T.~Wu
for discussions. B.-S. S. acknowledges support by the Swedish National
Research
Council under contract No. 8244-316, and by the Research Council of Norway
under the contract No. 420.95/004.


\def\pub#1#2#3#4#5{\bibitem{#2}#3,~#5.}

\end{document}